\documentstyle[prl,aps,multicol,epsfig]{revtex}
\begin{document}
\draft
\widetext

\newcommand{\be}{\begin{equation}}
\newcommand{\ee}{\end{equation}}
\newcommand{\ber}{\begin{eqnarray}}
\newcommand{\eer}{\end{eqnarray}}


\title{Macroscopic equations for pattern formation \\
in mixtures of microtubules and motors}

\author{Ha Youn Lee$^1$ and  Mehran Kardar$^{1,2}$}
\address{
$^1$Department of Physics, Massachusetts Institute of Technology,
Cambridge, Massachusetts 02139\\
$^2$Institute for Theoretical Physics, University of California, Santa Barbara, 
California 93106}

\maketitle

\begin{abstract}
\indent Inspired by patterns observed in mixtures of microtubules and
molecular motors, we propose continuum equations
for the evolution of motor density, and microtubule orientation.
The chief ingredients are the transport of motors
along tubules, and the alignment of tubules in the process.
The macroscopic equations lead to aster and vortex
patterns in qualitative agreement with experiments.
While the early stages of evolution of tubules are similar to
coarsening of spins following a quench, the rearrangement of motors
leads to arrested coarsening at low densities.
Even in one dimension, the equations exhibit a variety of interesting
behaviors, such as symmetry breaking, moving fronts, and motor
localization.

\end{abstract}
\widetext


\begin{multicols}{2}

\narrowtext
\section{Introduction}
\indent The emergence of complex patterns and correlated fluctuations
is a characteristic of many out of equilibrium 
situations\cite{Cross,Gollub}.
Living organisms provide many examples, ranging from the 
flocking of birds\cite{Vicsek,Toner},
the social organization of bacterial colonies\cite{Berg,Budrene,Ben-Jacob}, 
to internal reorganization of a cell during division\cite{Alberts,Hyman}.

 At the molecular level, motors and microtubules are frequently the
ingredients responsible for construction and motion.
Molecular motors are the proteins that convert chemical energy to mechanical
energy, and have been extensively studied\cite{Svoboda,Visscher,Fisher}.
Microtubules, consisting of a subunit protein called tubulin
provide the scaffolding for many cell constructs, as well as
``railways'' for transport of proteins\cite{Mitchison1,Dogterom}.
In particular, microtubules have a polarity that provides a direction
for the transport of motors.

 One of the many processes in which motors
and microtubules are involved is cell division.
They are indeed the main ingredients of mitotic spindles which serve to
move apart the duplicated chromosomes of eucaryotic cells\cite{Alberts,Hyman}.
To elucidate some of the physical mechanisms involved, 
several {\it in vitro} experiments  
on mixtures of microtubules and motors
have been carried out\cite{Urrutia,Nedelec,Surrey}.
Even these simple mixtures result in interesting patterns:
At an initial stage the microtubules form an aster
with their ``plus'' ends pointing towards a center.
The kinesin based beads move along the microtubules
towards this center.
In a confined geometry, 
the aster pattern is then destabilized, giving way to
a vortex in which the motors rotate around a center.
At larger scales and in unconfined geometries,
a variety of self-organized patterns are obtained
upon varying the motor concentration.
With increasing concentration,
an array of vortices, a mixture of asters and vortices,
a collection of asters, and bundles of microtubules emerge.

 The motivation of this paper is to describe some of the patterns
observed in these experiments.
There are in fact already several models of these phenomena
in the literature, starting with the simulations reported 
in the original paper on the patterns\cite{Nedelec}.
A two-dimensional model for the orientations
of microtubules was used to produce inhomogeneous 
stripe patterns\cite{Bassetti}.
Another recent model 
introduces a convection-diffusion
equation for motor density in the presence of the microtubule array.
This model results in a density profile of motors in asters
which decays as a power-law\cite{Nedelec2}.

 Our approach is to take a macroscopic perspective,
introducing two continuous fields
$m(\vec{r}, t)$ and $\vec{T} (\vec{r}, t)$
to describe the local motor density and tubule orientation,
respectively.
The central input to these equations is that the motor density is 
transported along the tubule direction, while the tubules are
in turn aligned by the motors.
In the {\em in vitro} experiments\cite{Nedelec},
the physical origin of the latter is that the motor
complexes can attach to two nearby tubules and their motion 
along the two provides a force that makes them parallel\cite{Kinesincomplex}.
In writing such continuum equations for densities,
we are following recent studies modeling the flocking
of birds\cite{Vicsek,Toner},
and the organization of growing bacteria\cite{Berg,Budrene,Ben-Jacob}.
Simulations of these equations
reproduce asters and vortices in agreement with experiment.
Analytical solutions also provide further insights on the patterns.
For example, we find that the motor density is much larger close
to the center of an aster, than in a vortex.
The resulting increased strain energy on the tubules provides
a driving force for asters to break off to form vortices,
in qualitative agreement with experiments.

 The global evolution of the tubule pattern is sensitive 
to the initial density of motors.
At high density, it is similar to coarsening of XY spins following a
quench from high temperatures\cite{Mondello,Pargellis,Bray,Rutenberg,Rojas}.
With periodic boundary conditions, the ultimate pattern is 
one of aligned tubules, with the motors going around in a uniform current.
Such a pattern is not possible with closed boundary conditions, which
typically lead to a single vortex in the center.
 At lower densities, fluctuations play a strong role,
and we observe a novel phenomenon of {\em arrested coarsening}, 
in which an inhomogeneous pattern of tubules freezes at some point in time.
This occurs because the transport of motors produces regions in which
the density of motors is very low (effectively zero). When such regions
percolate throughout the system,
no further rearrangement of tubules is possible.

 To better understand the dynamics of coarsening
and sizes of the arrested domains,
we also simulated the equations in one dimension.
Even in this case, the equations exhibit
a variety of interesting patterns that depend on the boundary
conditions:
(i) With periodic boundary conditions, there is a phase transition
between a state with a uniform current of motors running along tubules
aligned in one direction (at high motor densities);
and one in which there is a localized cluster of motors moving
at constant velocity around the system (at low densities).
Note that both patterns correspond to a broken symmetry (of the
two possible tubule orientations) in one dimension.
In contrast to the equilibrium Ising model, this symmetry breaking
appears to persist in the presence of noise (mimicking finite temperatures).
(ii) Reflecting boundary conditions lead to an oscillating
front sweeping back and forth across the system;
yet another solitonic solution to the equations.
(iii) Closed boundary conditions  give rise to initial coarsening,
and eventual freezing of the tubules into domains. 
The domain size for frozen tubules
depends on the value of the average motor density.
Unlike its two dimensional counterpart, the motor density continues to 
evolve after the tubules are frozen, and 
all motors are eventually localized in one cluster.

\section{Model}
\indent We introduce the local motor density $m(\vec{x},t)$, and 
the tubule orientation field $\vec{T}(\vec{x},t)$.
The conservation of motors leads to the continuity equation
\begin{equation}
{\partial m \over \partial t}
=-\nabla \cdot J_m, 
\end{equation}
where for the motor current, we shall assume the  form
\begin{equation}
J_m=-D ~\nabla m + A~m~\vec{T}.
\label{eq:motor0}  
\end{equation}
Here, $D$ is the diffusion constant for motors,
while $A$ is a coefficient describing their 
transport  along the tubule direction $\vec{T}$.
These are the lowest order terms in an expansion in $m$;
 higher order terms are expected and may become important at
high motor densities.
While tubules generally grow and fragment\cite{Dogterom};
their lengths can be stabilized by addition of taxol.
We model this preference towards a particular length by an
energy function, $-\alpha T^2/2+\beta T^4/4$,
similar to that used to describe spins in
the Landau-Ginzburg equation.
In the same spirit, we associate an energy cost of 
$K{(\nabla \vec{T})}^2/2$ with variations in the orientation of tubules.
Extremizing such an energy function leads to
\begin{equation}
{\partial \vec{T} \over \partial t}
=\alpha \vec{T} -\beta T^2 \vec{T} + \nabla \cdot (K \nabla \vec{T}). 
\label{eq:tubule0}  
\end{equation}
Since the alignment of tubules is entirely by motors,
we set $K=\gamma m$.
Of course, since this is a non-equilibrium process, there is no {\em a
priori} reason for the dynamics to originate from the minimization
of an energy functional. 
A more systematic approach is to include all terms allowed by symmetries. 
In this case the equation for $\partial\vec{T}/\partial t$ may
include terms $\gamma m\nabla^2\vec{T}$ and 
$\gamma'\nabla m\cdot\nabla\vec{T}$ with different coefficients.
For ease of description we shall start with the case $\gamma=\gamma'$, 
so that the dynamics is similar to minimizing
an energy locally proportional to 
$\gamma m(\vec{x},t) {(\nabla \vec{T})}^2/2$.

Finally, we can bring the equations into simpler form by rescaling to
\begin{eqnarray}
{\partial m \over \partial t}
&=& \nabla^2 m -  \nabla \cdot (m \vec{T}), 
\label{eq:motor} \\
{\partial \vec{T} \over \partial t}
&=&C \vec{T} \left(1-T^2 \right)+ \nabla \cdot ( m \nabla \vec{T} ),
\label{eq:tubule}  
\end{eqnarray}
where we now measure length in units of $\sqrt{\beta /\alpha} D/A$,
time in units of $\beta D / (\alpha A^2)$,
motor density in units of $D/\gamma$,
and the tubule density vector in units of $\sqrt{\alpha /\beta}$.
The remaining parameter  $C$ is given by $\beta D/ A^2$.

\section{Simulations}
\indent We perform numerical simulations on a two-dimensional $L\times L$ 
lattice, adapting the Crank--Nicholson scheme with the
ADI operator splitting method\cite{Press}.
The equations are discretized with spatial intervals of 
$\Delta x = \Delta y=1$,
and time intervals of $\Delta t=10^{-2}$.
At the edges of a finite system, we employ one of several possible
boundary conditions:
{\em Reflecting boundaries} have fixed inward pointing microtubules 
described by
\begin{equation}
\left . \vec{T} ~\right |_{\mbox {boundary}} = -\hat{n},
\end{equation}
where $\hat{n}$ is the normal outward vector at the boundary.
This discourages motors from approaching an edge.
By contrast, with {\em parallel boundary conditions} the microtubules
are tangential to the boundaries, while {\em closed boundary conditions}
place no restriction on the tubules. 
In all these cases, there is no current transporting motors outside
the system.
There is no restriction on the motor current when {\em periodic boundary
conditions} are applied, although again the total number of motors
is conserved.

\begin{figure}
\begin{center}
\epsfig{file=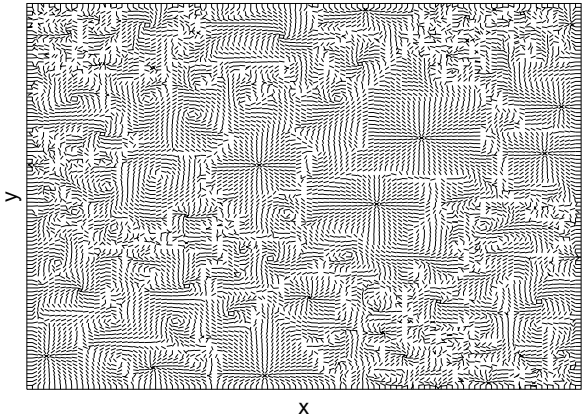,clip=,width=0.95\columnwidth}\
\end{center}
\vspace*{-1cm}
\caption[]{A lattice of vortices and asters for $m_0$=0.01 and $C$=100.
Initially, the size of the tubule is $10^{-3}$, and their directions
are random.
}
\label{fig:a-vlattice}
\end{figure}

\begin{figure}
\begin{center}
\epsfig{file=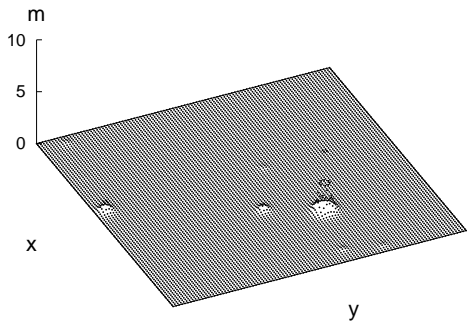,clip=,width=0.95\columnwidth}\
\end{center}
\vspace*{-1cm}
\caption[]{
The profile of motor density  corresponding to Fig.~\ref{fig:a-vlattice}.
The initial homogeneous density evolves to obtain
peaks at the centers of  asters and  vortices.
}
\label{fig:a-vlatticemotor}
\end{figure}

\begin{figure}
\begin{center}
\epsfig{file=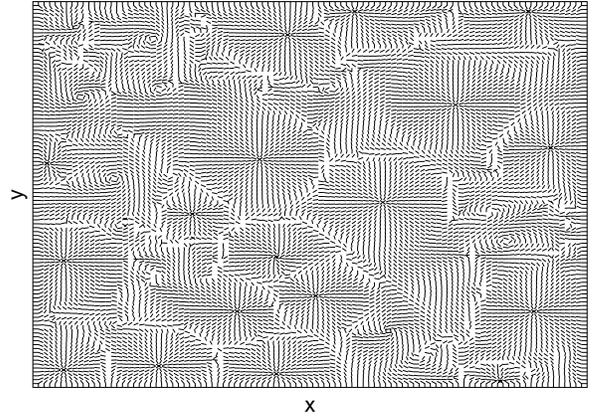,clip=,width=0.95\columnwidth}\
\end{center}
\vspace*{-0.5cm}
\caption[]{
A lattice of asters for $m_0=0.15$ and $C=100$.
}
\label{fig:alattice}
\end{figure}

\begin{figure}
\begin{center}
\epsfig{file=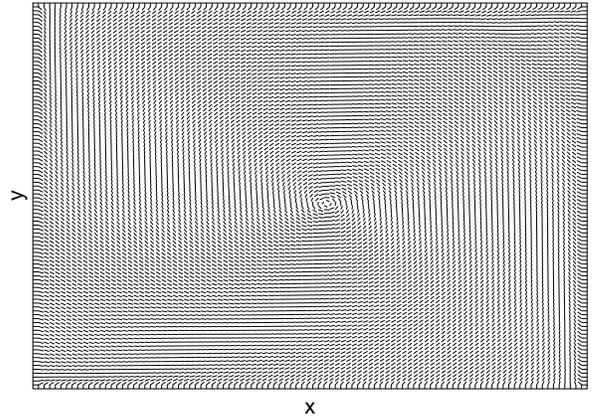,clip=,width=0.95\columnwidth}\
\end{center}
\vspace*{-0.5cm}
\caption[]{
A large vortex for $m_0=0.5$ and $C=100$.
At sufficiently high densities of motors, one or several vortices
are formed.
}
\label{fig:large vortex}
\end{figure}

We start with an initial condition in which the motor density is
uniformly set to $m_0$ at all points, while the tubule field
has magnitude $|\vec{T}_0|=10^{-3}$ and random orientations.
After a transient period,
the homogeneous configuration self organizes into patterns that
depend on the value of average motor density $m_0$,
as well as the growth constant $C$.
Figure \ref{fig:a-vlattice} shows a mixture of vortices and asters
which arises as the stationary pattern for $m_0=0.01$.
Both asters (tubules pointing inward) and vortices (tubules going
around) are clearly visible and randomly arranged throughout the system.
The  motor density now becomes inhomogeneous, with motors
accumulating in the centers of vortices 
and asters [Fig.~\ref{fig:a-vlatticemotor}].
The asters are more visible and dominant as the initial motor
density increases, as shown for $m_0=0.15$ in Fig.~\ref{fig:alattice}.
However, higher densities lead to a single 
vortex as in Fig.~\ref{fig:large vortex} for
$m_0=0.5$\cite{comment1}.

\vspace*{-1cm}

\begin{figure}
\begin{center}
\epsfig{file=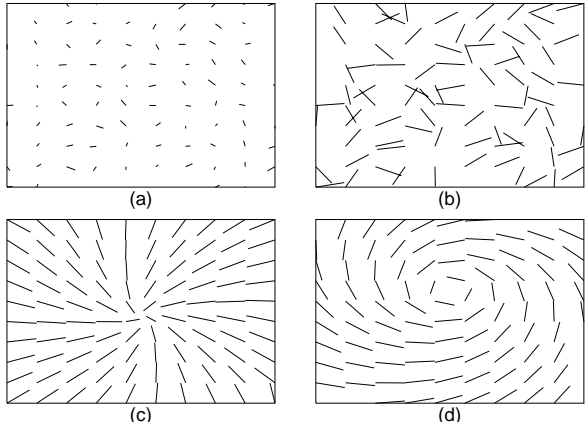,clip=,width=0.95\columnwidth}\
\end{center}
\vspace{0cm}
\caption[]{
The evolution of a vortex from an aster at times
$t=0.6$ (a), $t=1$ (b), $t=120$ (c), and $t=1200$ (d).
The random initial pattern of tubules first grows in
length, and then organizes as an aster.
Motors accumulate at the center of the aster, and
then circle around when it changes to a vortex.
}
\label{fig:dynamics}
\end{figure}

 The dynamics of formation of a large vortex 
from the random initial conditions is depicted in Fig.~\ref{fig:dynamics}.
This figure corresponds to reflecting boundary conditions, with
$C$=10, and $m_0$=0.15.
Since the initial tubule length is smaller than unity, the first stages
of evolution are the lengthening of tubules as depicted in
Figs.~\ref{fig:dynamics}(a) and \ref{fig:dynamics}(b), for times
$t=0.6$ and $t=1$, respectively.
During this stage the directions of the tubules do not change and
remain randomly distributed.
The next stage involves reorientations of the tubules.
Since a uniform alignment is incompatible with the boundary
conditions, 
an aster forms in the center as depicted in Fig.~\ref{fig:dynamics}(c),
for $t=120$.
Motors are now transported along the tubules and accumulate at
the center of the aster.
At longer times, the aster pattern gives way to a vortex
as in Fig.~\ref{fig:dynamics}(d) for $t=1200$.
The vortex pattern is stable, although its center may move around
depending on the choice of boundary conditions\cite{comment1}.
This dynamics is consistent with the experimental observation
that vortices form from the destabilization of asters\cite{Nedelec}.

\begin{figure}
\begin{center}
\epsfig{file=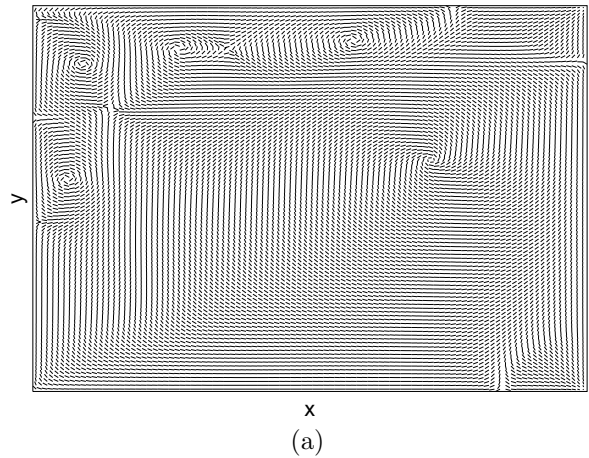,clip=,width=0.95\columnwidth}\\(a)\\
\epsfig{file=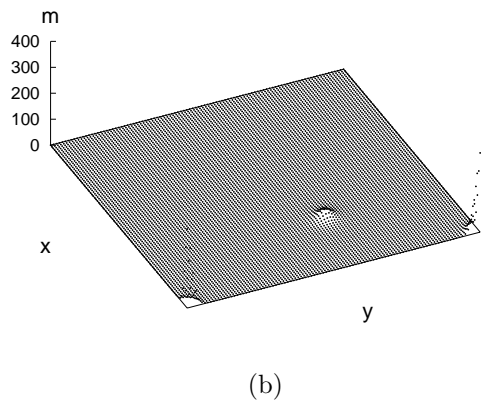,clip=,width=0.95\columnwidth}\\(b)\\
\end{center}
\caption[]{
The configuration of tubules for $m_0=0.5$, with closed boundary conditions (a);
and the corresponding profile of motor density (b).
Motors pile up at several points on the boundary, 
in addition to the interior.
}
\label{fig:OBmotorconfig}
\end{figure}

 To model pattern formation in larger systems in which boundary effects
are less important, we also performed simulations with closed
boundary conditions.
This change in the boundary condition does not qualitatively
alter the pattern formation process.
At  low motor densities, we still observe a mixture of vortices
and asters, followed by a collection of asters as  $m_0$ is increased.
However, at large motor density, after formation of a large vortex
at the center, motors also pile up at several points on the
boundary, as indicated in Fig.~\ref{fig:OBmotorconfig}.


Equations~(\ref{eq:motor}) and (\ref{eq:tubule}) are  deterministic;
the only stochasticity appearing through the initial conditions.
However, randomness and noise are certainly present in the experimental
situations. In particular, microtubules are known to constantly grow
and shrink though a dynamic instability\cite{Mitchison1,Dogterom},
while asters are still observed under such conditions\cite{Mitchison2}.
To make sure that the patterns observed in our simulations survive
the addition of noise, we also 
introduced a stochastic noise in the tubule evolution equation.
We observed that the self-organized patterns are stable at small noise, 
but that sufficiently large noise causes a phase transition 
to homogeneous mixtures (bundles of microtubules in a uniform motor density).
Alternatively, we observe that for a fixed amount of noise,
patterns are destroyed at motor densities lower than a critical $m_c$,
but are qualitatively unchanged otherwise.
(The value of $m_c$ is 0.005 if the noise distributed uniformly between
$-1$ and $1$.)

\section{Aster and vortex solutions}
\begin{figure}
\begin{center}
\epsfig{file=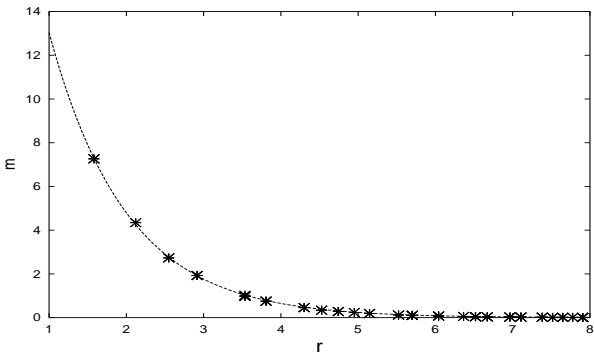,clip=,height=4.7cm,width=0.95\columnwidth}\
\end{center}
\caption[]{
The points represent the profile of  motor density in an aster pattern
for $m_0=0.1$ and $C=100$.
The dashed line is a least-square fit to the exponential decay
 $35.4 \exp(-r)$.
}
\label{fig:aster expodecay}
\end{figure}

\indent We can easily find analytical solutions to 
Eqs.~(\ref{eq:motor}) and (\ref{eq:tubule})
that describe the motor density in the center of an aster or vortex.
To this end, we look for  stationary solutions, 
$\partial_t~ m=\partial_t~ \vec{T}=0$,
with radial symmetry.
In an aster the tubules are directed towards the center
and $\vec{T}=-\hat{r}$, where $\hat{r}$ is the unit radial vector.
Balancing the diffusive current $-\partial_r m$, with that transported 
along tubules gives an exponential form
\begin{equation}
m_{\rm aster}(r)=m(0) \exp(-r). \label{eq:expodecay} 
\end{equation}
This exponential profile is indeed verified by the simulations 
as depicted in Fig.~\ref{fig:aster expodecay}.

\begin{figure}
\begin{center}
\epsfig{file=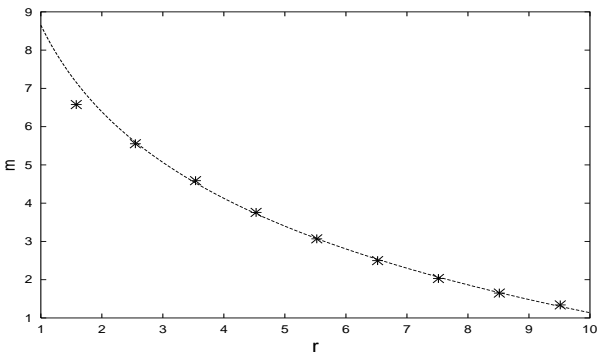,clip=,height=4.7cm,width=0.95\columnwidth}\
\end{center}
\caption[]{
The simulated profile of the motor density of a vortex
for $m_0=0.2$ and $C=1$.
The dashed line is the least-square fit to the form, $-3.3\log(r/14.2)$.
}
\label{fig:vortex logdecay}
\end{figure}

 The tubules go around the center of a vortex, and $\vec{T}=\hat{\theta}$,
where $\hat{\theta}$ indicates the tangential unit vector.
Motors are then transported in a uniform circular current by the tubules.
To ensure that there is no radial current of motors we need
$\partial_r (r \partial_r m)=0$,
whose solution is
\begin{equation}
m_{\rm vortex}(r)=-M \log\left({ r / R}\right), \label{eq:logdecay}
\end{equation}
where $R$ is a long distance cut-off, of the order of the vortex size.
A logarithmic fit to the simulated vortex motor density profile is 
shown in Fig.~\ref{fig:vortex logdecay}.

The profile of motors in quasi two dimensional asters has
been studied theoretically and experimentally in Ref.~\cite{Nedelec2}.
In this study the tubules can grow to form quite long tracks,
and their  density also falls off away from the aster center.
These differences in tubule behavior (as opposed to our case where
tubules have uniform density and length) lead to predicted
power law decays of the motor density profile.

 The aster and vortex configurations of tubules are related to
topological defects in the XY model.
However,  they are equivalent defects in the XY model as one can
be deformed into the other through a $90^\circ$ rotation.
The presence and rearrangement of motors in our problem 
breaks this symmetry and the two configurations become inequivalent.
In particular, the two defects have very different static
energies $E=\int d^2 r m(r) {(\nabla T)}^2/2$.
In the aster, the motors are concentrated close to the center
leading to a high strain energy.
By contrast, the motor density in a vortex in more uniform.
Consequently, for the same number of motors, a large aster has
much higher energy than a large vortex.
Since the dynamics tends to minimize this energy,
we have an explanation for why asters give way to the more stable
vortices.
Presumably, finite size effects in smaller asters, of the order
of the decay length implicit in Eq.~(\ref{eq:expodecay}), account
for their stability at small motor densities as in Fig.~\ref{fig:alattice}.

\section{Arrested Coarsening}
\indent If the motor density is maintained at a uniform and fixed value,
the dynamics of the tubules is identical to the coarsening of an
XY system following a quench from high temperatures.
This problem has been extensively 
studied\cite{Mondello,Pargellis,Bray,Rutenberg,Rojas}
and (up to logarithmic corrections) 
the typical length scale of the pattern coarsens
as $\xi \sim t^{1/2}$.
In our case, the motors rearrange themselves 
in the landscape of the tubules and the coarsening scenario is
modified when the motor density fluctuations become significant.

\begin{figure}
\begin{center}
\epsfig{file=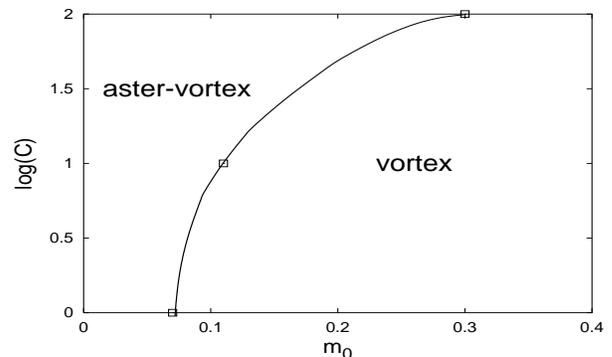,clip=,height=4.9cm,width=0.95\columnwidth}\
\end{center}
\vspace{-.3cm}
\caption[]{
The phase boundary in the plane of $m_0$ and $\log C$.
To the right of the boundary the pattern coarsens to a large vortex.
The arrested coarsening on the left side leads to frozen asters and
vortices, with the aster pattern becoming more
dominant as $m_0$ increases.
}
\label{fig:phase}
\end{figure}

 At high motor densities there are enough motors left over after formation
of asters and vortices to cause further rearrangements of the tubules,
and coarsening continues towards the final pattern consistent with
the boundary conditions.
However, at low densities the motors quickly migrate to the 
centers of asters  and vortices.
The little motor density left in the regions in between defects 
may then be too small to cause further realignment of tubules 
which became frozen. We call this phenomenon {\it arrested coarsening}
of tubules.
 The limiting value of $m_0$ for the onset of arrested coarsening
in fact depends on the growth constant of tubules $C$,
as indicated by the ``phase diagram" sketched in Fig.~\ref{fig:phase}.
 
\section{One dimension}
\indent To further understand the patterns, we examine the equations
in one dimension, where more detailed simulations are possible.
In particular, we consider scalar fields $m(x,t)$ and $T(x,t)$
evolving as
\begin{eqnarray}
\partial_t m &=& {\partial_x}^2 m - \partial_x (T m) , \label{eq:1dmotor}\\
\partial_t T &=& C (T-T^3) + \partial_x (m \partial_x T) \label{eq:1dtubule}.
\end{eqnarray}
Such equations may be appropriate to describe
the movement of myosin motors along quasi-one dimensional bundles of 
actin filaments which occurs in muscle contraction, and may be responsible
for other types  of cell motion\cite{Alberts}.
Bundles of actin molecules can be formed {\em in vitro} in mixtures
with other inert polymers.
These bundles consist of random mixtures of actin molecules oriented in
the two possible direction.
However, in the presence of myosin motors and ATP, there is a sorting
of polarity\cite{Takiguchi,Nakazawa} and 
active contraction of polar filaments\cite{Kruse}.
While the bundles are usually linear with open ends, then sometimes
assemble into a ring\cite{Takiguchi}.
It may also be possible to artifically confine other mixtures
of tubules and motors in
quasi-one dimensional containers with various boundary conditions.
In fact, we observe that the solutions to Eqs.~(\ref{eq:1dmotor})
and (\ref{eq:1dtubule}) are quite sensitive to boundary conditions.
Specifically, simulations show the following results:

{\em (i) Periodic boundary conditions} correspond to placing the
system on a closed loop.
We observe two types of symmetry breaking, depending on the 
initial density of motors.
If $m_0$ is larger than a critical value of $m_c$ ($m_c\approx 0.04$ for
$C=1$),
the tubule pattern coarsens until the tubules are all
aligned in one direction $(T=+1$ or $-1)$. 
This is accompanied by a uniform current of motors 
that goes around the system.
This symmetry breaking is similar to that of an Ising model.
However, unlike the equilibrium Ising model, the broken symmetry
survives in the presence of random noise 
(simulating finite temperatures) in Eq.~(\ref{eq:1dtubule}). 
Any domains of opposite spin formed due to randomness are annealed by
a rush of motors to the domain walls.
The symmetry breaking in the presence of noise is  due to advection
term in Eq.~(\ref{eq:1dmotor}), and active transportation of motors
along the tubules. Indeed we checked that when the advection term is
removed, the presence of noise in Eq.~(\ref{eq:1dtubule}) leads to
finite domains.

\begin{figure}
\begin{center}
\epsfig{file=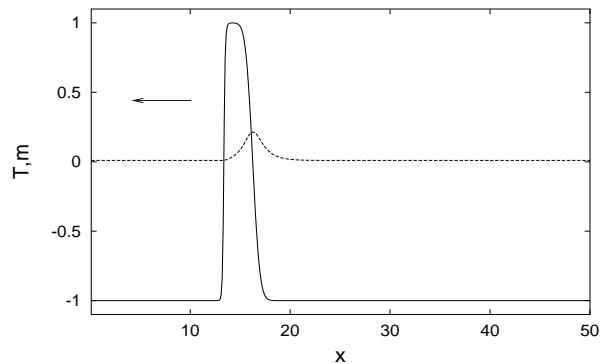,height=5cm,width=0.95\columnwidth}\
\end{center}
\caption[]{
The profiles of  motor density (dashed line) and  tubule orientation
(solid line) in one dimension with {\em periodic boundary conditions},
for $m_0=0.02$ and $C=1$.
The whole pattern moves as indicated by the arrow, with a fixed speed.
}
\label{fig:1dperi}
\end{figure}

If $m_0$ is smaller than $m_c$, a novel final state emerges in
which the motors gather together in a cluster that moves around
the loop with a constant velocity.
The tubules are again ordered in one direction, except near the
cluster, where they briefly take the opposite alignment. 
Figure \ref{fig:1dperi} shows that the configurations of the tubules
and the profile of the motor density for $m_0=0.02 < m_c$.
We have verified by direct numerical integration that 
Eqs.~(\ref{eq:1dmotor}) and (\ref{eq:1dtubule}) do indeed support
such a solitonic solution. 
However, the mathematical details are not in the spirit of this article,
and will be presented elsewhere.

\begin{figure}
\begin{center}
\epsfig{file=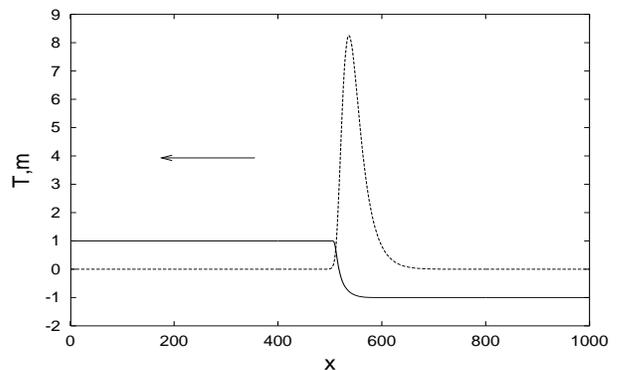,height=5cm,width=0.95\columnwidth}\
\end{center}
\vspace*{-0.3cm}
\caption[]{
The profiles of  motor density (dashed line) and  tubule orientation
(solid line) in one dimension with {\em reflecting boundary conditions},
for $m_0=0.4$ and $C=1$.
The cluster of motors moves as indicated by the arrow,
but its direction is reversed at each boundary, as the profile
oscillates back and forth.
}
\label{fig:reflecting}
\end{figure}

{\it (ii) Reflecting boundary conditions} were imposed by requiring the 
tubules at the edges to point inward, i.e. $[T(0)=+1$ and $T(L)=-1]$.
There is an initial coarsening period in which domains of $+1$ and $-1$ grow
inward from the respective edges.
However, in the final pattern the boundary between the $+1$ and $-1$ domains
is not stationary, but sweeps back and forth across the system!
The motors are again concentrated at the interface,
with a solitonic profile as indicated in Fig.~\ref{fig:reflecting}.

\begin{figure}
\begin{center}
\epsfig{file=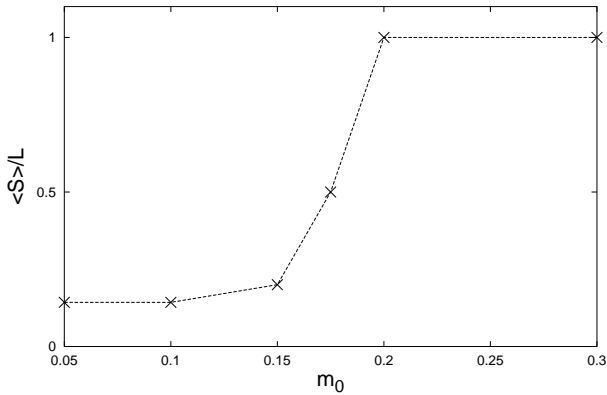,clip=,width=0.95\columnwidth}\
\end{center}
\caption[]{
The average tubule domain size in a one dimensional system of length $L$
with closed boundary conditions, as a function of average motor density
$m_0$ for $C=1$.
}
\label{fig:cb2}
\end{figure}

\begin{figure}
\begin{center}
\epsfig{file=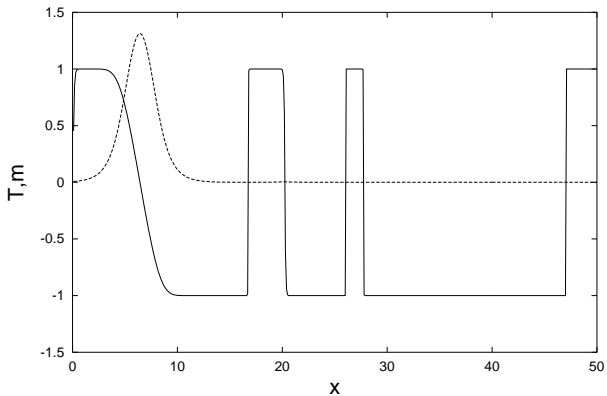,clip=,width=0.95\columnwidth}\
\end{center}
\vspace{0.5cm}
\caption[]{
The profiles of tubule orientation and motor density
(solid and dashed lines, respectively) for $m_0=0.1$ and $C=1$,
in a one dimensional system with closed boundaries.
}
\label{fig:cb1}
\end{figure}

{\it (iii) Closed boundary conditions} were also applied, with
no restrictions on the value of $T$ at the edges, but setting the
outward motor current to zero.
Above a critical motor density,  coarsening of tubules
proceeds to a single domain of the size of the system, $L$.
Following the tubules, the motors then pile up at one end of the system.
The low density behavior in this case is similar to the arrested
coarsening observed in the two dimensional case: 
The initial growth of $+1$ and $-1$ domains is stopped at some point 
due to the local absence of motors necessary for continuing realignments.
Figure \ref{fig:cb2} shows the average domain size,
as a function of the average motor density.
At the point when tubule evolution is stopped, 
the motor density  has peaks at $(+,-)$ domain boundaries.
However, as time goes on, there is a slower
ripening process in which the motors 
gradually diffuse against the unfavorable domains, and eventually
aggregate at one point in the system, as in Fig.~\ref{fig:cb1}.
(It is not clear if this process is truly absent in two dimension,
or merely takes too long to observe in simulations.
There are similarities to the localization problem in which 2 is a critical
dimension.)

There are similarities in the above behaviors to patterns and phase
transitions in other one dimensional non-equilibrium processes.
A localization transition is observed in a population with
diffusion, drift, and reproduction in Refs.~\cite{Nelson1,Nelson2,Nelson3}.
This model of a bacterial colony  living on an oasis in the presence 
of randomly fluctuating wind, shows extended, localized, mixed states,  
as well as extinction, as
a function of the average growth rate and convection velocity.
In our model, the tubules (in their disordered state) provide a
similar source of random advection. 
Ordered states are also observed in a lattice model with two species
of particles in which the mobility of one species depends on the density
of the other\cite{Rama1,Rama2}.
This model exhibits three phases: One with strong phase, 
a fluctuation-dominated phase, and another with uniform overall density.
The couplings in our system between motors and tubules have similarities
to the latter model, although the enforced conservation laws are
different (there is no conservation of tubule orientations).

\section{Discussion}
\indent Given how little input is used to construct the macroscopic 
equations (tubules transport motors and are aligned in the process),
it is encouraging that many features of the experimental patterns
are reproduced. As in the experiments, we  observe 
arrays of asters and vortices, 
and large single vortices (formed from the break up of asters). 
However, at high densities of motors, the experiments lead to
irregularly arranged bundles of tubules, a feature
not present in our model.
The absence of tubule bundles indicates the limitations of the macroscopic
approach. To reproduce the observed sequence of patterns, more physical input
into the equations is necessary. For example, a potential modification
is to include non-linear terms in the transported motor current 
to model the inter-particle interactions at high densities.

\begin{figure}
\begin{center}
\epsfig{file=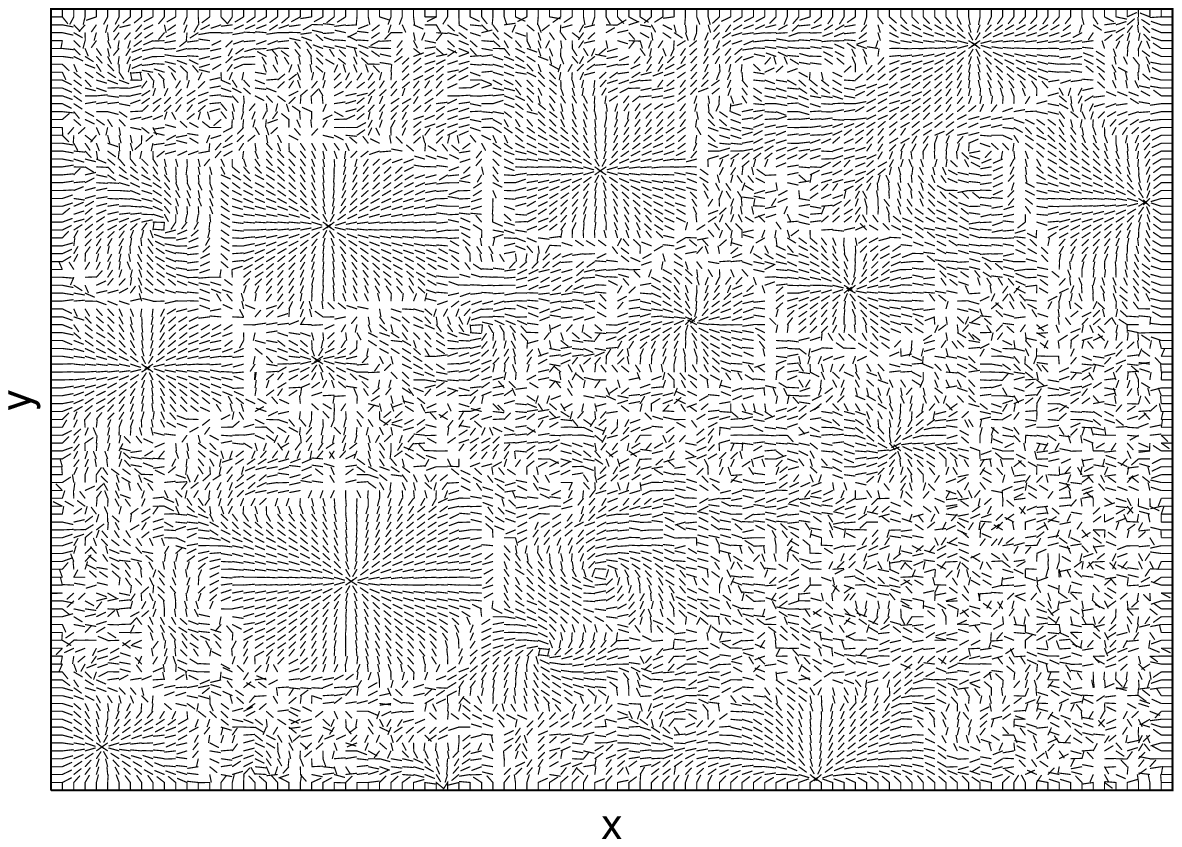,height=5cm,width=0.95\columnwidth}\\(a)\\
\epsfig{file=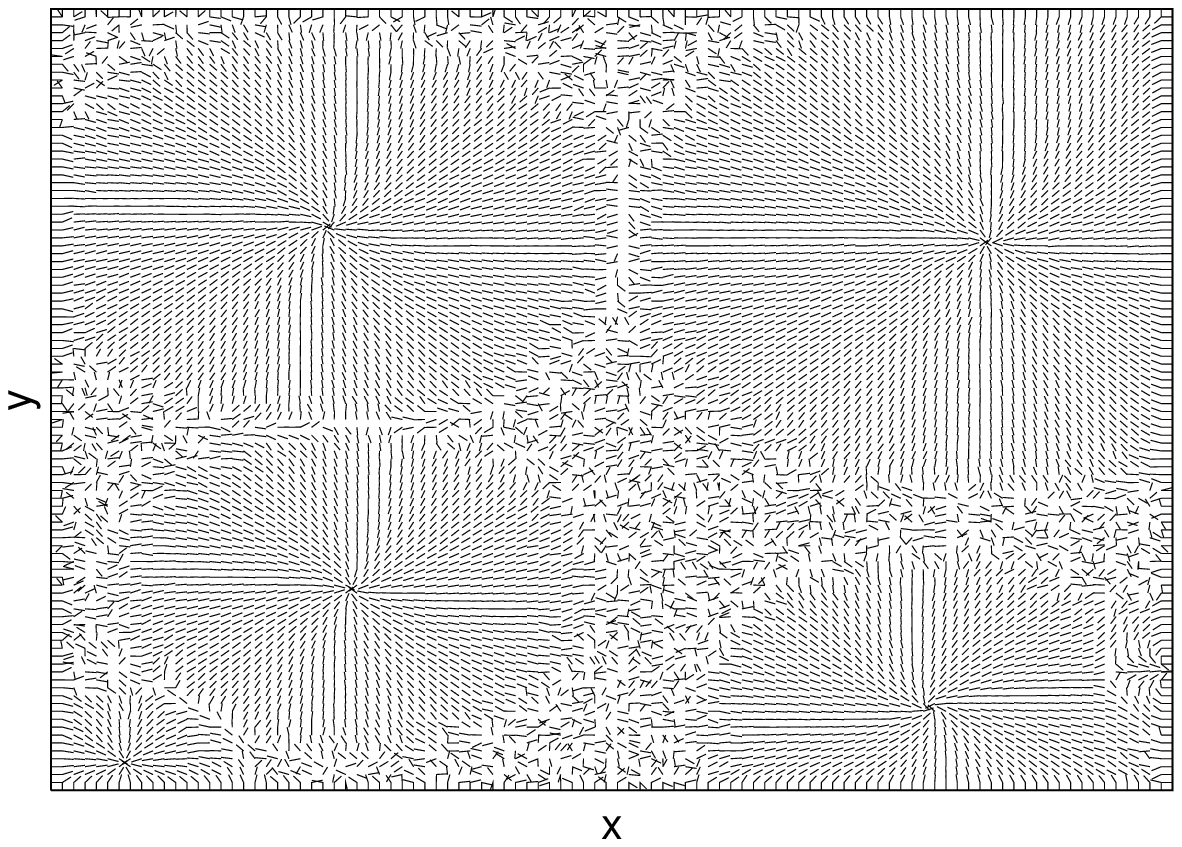,height=5cm,width=0.95\columnwidth}\\(b)\\
\epsfig{file=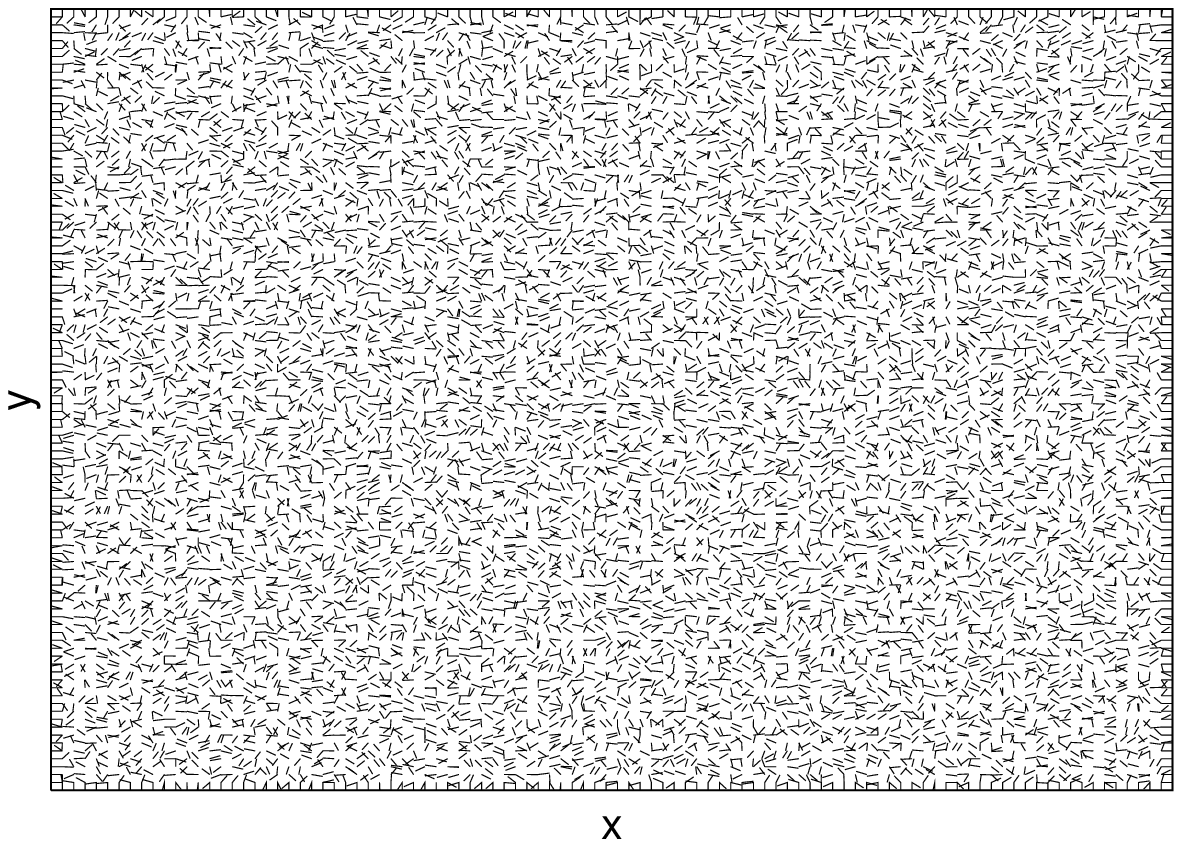,height=5cm,width=0.95\columnwidth}\\(c)
\caption[]{
The sequence of patterns obtained upon inclusion of saturation effects
for $m_0=0.05$ (a), $m_0=1$ (b), and $m_0=20$ (c).
}
\label{fig:CUT}
\end{center}
\end{figure}

In models of highway traffic\cite{Kerner}, the current actually
diminishes at high density,
and can be approximated by
\begin{equation}
J[m]=-\nabla m + \vec{T}~ m~ e^{-m/m_{\rm max}},
\end{equation}
where $m_{\rm max}$ is a characteristic density for saturation of current.
Since the alignment of tubules is intimately connected to the motor current,
similar modifications should appear in Eq.~(\ref{eq:tubule}).
As discussed earlier, because of the non-equilibrium nature 
of the process, the equation for $\partial_t \vec{T}$ does not
have to be obtained from the minimization of an energy density.
In fact, to describe saturation effects on alignment of tubules,
we modified Eq.~(\ref{eq:tubule}) to
\begin{equation}
\partial_t \vec{T} = C \vec{T} \left(1-T^2\right) 
+ \nabla m \cdot \nabla \vec{T}
+ m e^{-m/m_{\rm max}} \nabla^2 \vec{T} + \eta,
\end{equation}
where $\eta$ is a stochastic noise.
The sequence of tubule patterns obtained from these modified
equations upon increasing
initial motor density is depicted in Fig.~\ref{fig:CUT}.
We observe that the single vortex pattern becomes unstable at high
densities, and we may regard 
the high density phase as representing bundles of tubules.
We also find that the patterns of vortices and asters 
are stable in the presence of noise in the following sense:  
If we average the whole configuration of tubules over time, 
the intermediate regions between defects are washed away while the 
defects survive.

 The rich variety of behaviors observed in these simple equations
are due to their nonequilibrium character.
In the realm of equilibrium, for example, one dimensional systems 
at non-zero temperature are featureless and disordered. 
Clearly non-equilibrium effects can lead to symmetry breaking, 
and a rich interplay behaviors sensitive to boundaries. 
Biological systems can take advantage of such phenomena, 
and should indeed provide many interesting patterns in need 
of explanation.

We wish to thank F. J. N\'{e}d\'{e}lec and  P. De Wulf
for helpful discussions.
H.Y.L. is supported by the Korea Science and Engineering Foundation
through a fellowship.
MK acknowledges the support of 
NSF grants DMR-98-05833 (at MIT) and PHY99-07949 (at ITP).

\end{multicols}


\begin{thebibliography}{7}
\bibitem{Kinesincomplex} In Ref.\cite{Nedelec}, an artificial molecular
construct consisting of four  two-headed kinesin
molecules linked by biotin-streptavidin was used.
These molecules can simultaneously bind to two microtubules.
However, it is not known if binding of microtubules by two-headed
kinesin molecules occurs inside the cell, or if other mechanisms
are responsible for their alignment.
\bibitem{comment1} With  boundary conditions of inward pointing tubules,
the center of the single vortex moves around the center of the system.
However, with tubules aligned parallel to the
 boundary, the pattern of the final vortex is stationary.
\end{thebibliography}
\end{document}